\documentclass[twocolumn,showpacs,preprintnumbers,amsmath,amssymb,superscriptaddress,prl]{revtex4-1}

\usepackage{graphicx}
\usepackage{dcolumn}
\usepackage{bm}
\usepackage{graphics}
\usepackage{subfigure}
\usepackage{graphicx}
\usepackage{epsfig}
\usepackage{epstopdf}
\usepackage{color}

\begin{document}

\title{Plasmon signature in Dirac-Weyl liquids}

\author{Johannes Hofmann}

\author{S. Das Sarma}

\affiliation{Condensed Matter Theory Center and Joint Quantum Institute, Department of Physics, University of Maryland, College Park, Maryland 20742-4111 USA}

\date{\today}

\begin{abstract}
We consider theoretically as a function of temperature the plasmon mode arising in three-dimensional Dirac liquids, i.e., systems with linear chiral relativistic single-particle dispersion, within the random phase approximation. We find that whereas no plasmon mode exists in the intrinsic (undoped) system at zero temperature, there is a well-defined finite-temperature plasmon with superlinear temperature dependence, rendering the plasmon dispersion widely tunable with temperature. The plasmon dispersion contains a logarithmic correction due to the ultraviolet-logarithmic renormalization of the electron charge, manifesting a fundamental many-body interaction effect as in quantum electrodynamics. The plasmon dispersion of the extrinsic (doped) system displays a minimum at finite temperature before it crosses over to the superlinear intrinsic behavior at higher temperature, implying that the high-temperature plasmon is a universal feature of Dirac liquids irrespective of doping. This striking characteristic temperature dependence of intrinsic Dirac plasmons along with the logarithmic renormalization is a unique manifestation of the three-dimensional relativistic Dirac nature of quasiparticle excitations and serves as an experimentally observable signature of three-dimensional Dirac materials.
\end{abstract}

\maketitle

In Dirac semimetals the valence and conduction bands touch only in isolated points of the Brillouin zone with a Fermi energy that is tuned close to these band touching points, giving rise to particle excitations with an effective linear relativistic band dispersion $\varepsilon_\pm(k) = \pm \hbar v_F k$. Much of the interest in these novel materials stems from the topological properties of Weyl semimetals which are manifest in anomalous transport properties~\cite{yang11,zyuzin12} and Fermi arc surface states~\cite{wan11}. Recent hallmark experiments have realized these materials in Cd${}_3$As${}_2$, Na${}_3$Bi, and TaAs and have demonstrated the Dirac cone structure in ARPES measurements~\cite{liu14,neupane14,borisenko14,liu14b,jeon14,xu15b} and detected topological surface states in Na${}_3$Bi and TaAs~\cite{xu15,xu15b}. These initial experiments, which serve to verify the expected band structure, motivate the search for definitive experimental signatures of Dirac materials which establish unique observable effects distinct from standard electron gases and normal metals. In this Rapid Communication, we show that the collective charge oscillation of a Dirac liquid, the plasmon, is widely tunable as a superlinear function of temperature (with $\sim - T/\ln T$), which allows us to extract high-precision information on the Dirac material. Moreover, the dispersion reveals a subtle logarithmic electron charge renormalization effect which is an additional specific signature of a three-dimensional Dirac system. Note that previous work exclusively studies the plasmon mode at zero temperature~\cite{lv13,son13,liu13,panfilov14,zhou14}. In this Rapid Communication, we establish a different mechanism for finite-temperature plasmons which arises due to the thermal excitation of electrons from the valence to the conduction band. It turns out that this interband mechanism overwhelms the low-temperature canonical intraband mechanism at sufficiently large temperature, resulting in a universal behavior that is independent of initial detuning.

The collective plasmon mode of an electron system is defined as the complex pole $z=\omega(q)-i\gamma(q)$ of the dielectric function $\varepsilon(z,{\bf q})$; i.e., it solves
\begin{align}
\varepsilon(z,{\bf q}) = 1 - V_{\bf q} \Pi(z,{\bf q}) &= 0 , \label{eq:defpole}
\end{align}
where $\Pi(z, {\bf q})$ denotes the polarizability and $V_{\bf q}$ is the Coulomb interaction $V_{\bf q} = 4 \pi e^2/\kappa {\bf q}^2$ with $e$ being the electron charge and $\kappa$ the effective dielectric constant of the medium. The dimensionless interaction strength (the Dirac fine structure constant) is defined by the ratio of interaction and kinetic energy
\begin{align}
\alpha &= \frac{e^2}{\hbar v_F \kappa} .
\end{align}
Here, because of linear band dispersion, we have the special situation that both kinetic and potential energy scale as $n^{1/3}$ with carrier density, implying that $\alpha$ is density independent. This manifests a scale invariance of the theory on the classical level. By contrast, the nonrelativistic system with quadratic dispersion $\hbar^2 k^2/2m$ is generically scale-invariant only for $1/r^2$ long-range interactions (or, for short-range interactions, by fine-tuning to the unitary limit of infinite scattering length), implying that for the nonrelativistic Coulomb interactions the effective interaction strength $r_s \sim n^{-2/3}$ can be changed by altering the density.

The form of the plasmon dispersion both at zero and finite temperature is determined by dimensional analysis: at zero temperature and finite density, the plasmon dispersion $\omega_p \sim  \varepsilon_F \sim n^{1/3}$, where $\varepsilon_F$ is the Fermi energy and $n$ is the doping-induced carrier density, approaches a constant as is characteristic for three-dimensional systems. It should be noted that in contrast to the electron liquid the plasmon frequency $\omega_p$ of the Dirac liquid is an entirely quantum mechanical quantity~\cite{dassarma09} with $\hbar$ appearing explicitly in the leading-order low-momentum term of the plasmon dispersion, $\omega_p \sim 1/\sqrt{\hbar}$ [whereas the electron gas with $\omega_p = \sqrt{4\pi n e^2/\kappa m}$ receives quantum corrections only in higher order ${\it O}(q^2/\hbar^2)$]. At zero temperature and zero density (the intrinsic case), by dimensional analysis we expect the plasmon dispersion to vanish. At finite temperature, on the other hand, thermal excitations of particles from the valence to the conduction band are not gapped and a solution of Eq.~\eqref{eq:defpole} exists even at zero doping. Here, the low momentum plasmon frequency should be proportional to temperature, $\omega_p \sim  T$. It turns out that this simple scaling analysis is not quite correct and the full solution predicts $\omega_p \sim  - T/\ln T$: both plasmon dispersion and damping receive a quantum correction due to the renormalization of the Coulomb interaction, a field-theoretical effect in interacting many-body systems which does not possess an analog in standard electron gases with parabolic band dispersion or Dirac materials in other dimensions~\cite{dassarma13}.

In this Rapid Communication, we formalize this discussion by a full random phase approximation (RPA) calculation of the plasmon dispersion at finite temperature and density. We compute analytically the plasmon dispersion and damping, finding a characteristic superlinear temperature dependence which includes a logarithmic renormalization correction for the intrinsic Dirac liquid. For the extrinsic doped system, our calculations indicate a crossover from a low-temperature extrinsic regime to a high-temperature intrinsic regime where the plasmon behaves as in the undoped system. We also complement all our analytical findings with full numerical calculations of the plasmon properties. In addition to the plasmon dispersion, we present numerical results for the dielectric function as a function of momentum, frequency, and temperature for the Dirac-Weyl system.

Within RPA, we compute the plasmon mode in Eq.~\eqref{eq:defpole} using the noninteracting polarizability
\begin{align}
\Pi(z,{\bf q}) &= 
\frac{g}{V} \sum_{{\bf k}ss'} \, \frac{f(\varepsilon_s({\bf k})) - f(\varepsilon_{s'}({\bf k}'))}{\hbar z + \varepsilon_s({\bf k}) - \varepsilon_{s'}({\bf k}')} \, F_{ss'}({\bf k}, {\bf k}') . \label{eq:poldef}
\end{align}
Here, ${\bf k}' = {\bf k} + {\bf q}$, $f(\varepsilon_s({\bf k})) = n_F(\varepsilon_s({\bf k}) - \mu)$ is the Fermi-Dirac distribution, $\mu$ the chemical potential, and $F_{ss'}({\bf k}, {\bf k}') = (1+ss' \cos \theta)/2$ is the overlap of two wave functions with energy $\varepsilon_s(k)$ and $\varepsilon_{s'}(k')$ which can be evaluated in terms of the relative angle $\theta$ between ${\bf k}$ and ${\bf k}'$. $g$ accounts for a degeneracy of Dirac cones in the material. Equation~\eqref{eq:poldef} diverges logarithmically in momentum~\cite{abrikosov70,goswami11,hofmann14}, and we introduce a scale $\Lambda$ beyond which we cut off the linear Dirac dispersion. Throughout this Rapid Communication, we set $k_B=1$ but retain the dependence on $\hbar$ where appropriate.

We can split the polarizability into a zero-temperature intrinsic contribution which encompasses the complete cutoff dependence and a correction due to finite temperature and doping. At zero temperature, the intrinsic polarizability reads
\begin{align}
\Pi(\omega , {\bf q}) &= - \frac{g q^2}{24 \pi^2 \hbar v_F} \ln \biggl| \dfrac{\Lambda^2}{\omega^2/v_F^2 - q^2} \biggr| \nonumber \\
&- \frac{i g q^2}{24 \pi \hbar v_F} \Theta(\omega- v_F q) \label{eq:polint} .
\end{align}
Substituting this in Eq.~\eqref{eq:defpole}, we notice that the expression does not involve $\alpha$ and $\Lambda$ separately but combined in the form of a single dimensionful scale $\Lambda_L = \Lambda e^{3 \pi/g\alpha}$, the Landau pole~\cite{hofmann14}. In typical Dirac materials with $\kappa \gg 1$, we find the following hierarchy of scales:
\begin{align}
k_F \ll \Lambda \ll \Lambda_L .
\end{align}
Note that even a weak screening increases the Landau pole scale exponentially compared to the cutoff $\Lambda$. This appearance of a dimensionful scale is known as dimensional transmutation and signifies the breakdown of classical scale invariance. This so called scale anomaly, where a system that is scale-invariant on a classical level acquires a dimensionful scale through the process of renormalization, is present in other condensed matter or atomic systems as well: the most pivotal example is the standard BCS theory with attractive contact interaction of strength $g$. Here, the renormalization of the dimensionless coupling implies that the combination $\Delta = \omega_D e^{1/VN(0)g}$ ($\omega_D$ being the Debye frequency, i.e., the cutoff scale of the theory) is an observable quantity independent of the renormalization group scale. Moreover, in two-dimensional atomic Fermi gases, the breaking of scale invariance is conjectured to cause a shift in the collective breathing mode~\cite{vogt12,hofmann12} and to induce logarithmic scaling violations in the high-frequency tail of radiofrequency absorption spectra~\cite{langmack12} and other response functions~\cite{hofmann11}. It should be noted, however, that in the above examples the scale arising due to dimensional transmutation is an inherently low-energy scale (the superconducting gap or the dimer bound state energy, respectively), whereas in the Dirac materials discussed in this work the Landau scale is larger than the cutoff $\Lambda$. We remark that this scale anomaly is inherently a nonperturbative effect involving an essential singularity in the coupling constant $\alpha$. Note that the cutoff $\Lambda$ corresponds to the momentum scale beyond which the dispersion is no longer linear. For Weyl materials with Weyl pairs in close proximity in the Brillouin zone (such as the pyrochlore iridates or TaAs~\cite{huang15,xu15b}), the Landau pole could be sufficiently small to ensure large logarithmic scaling corrections.

\begin{figure}[t]
\scalebox{0.6}{\includegraphics{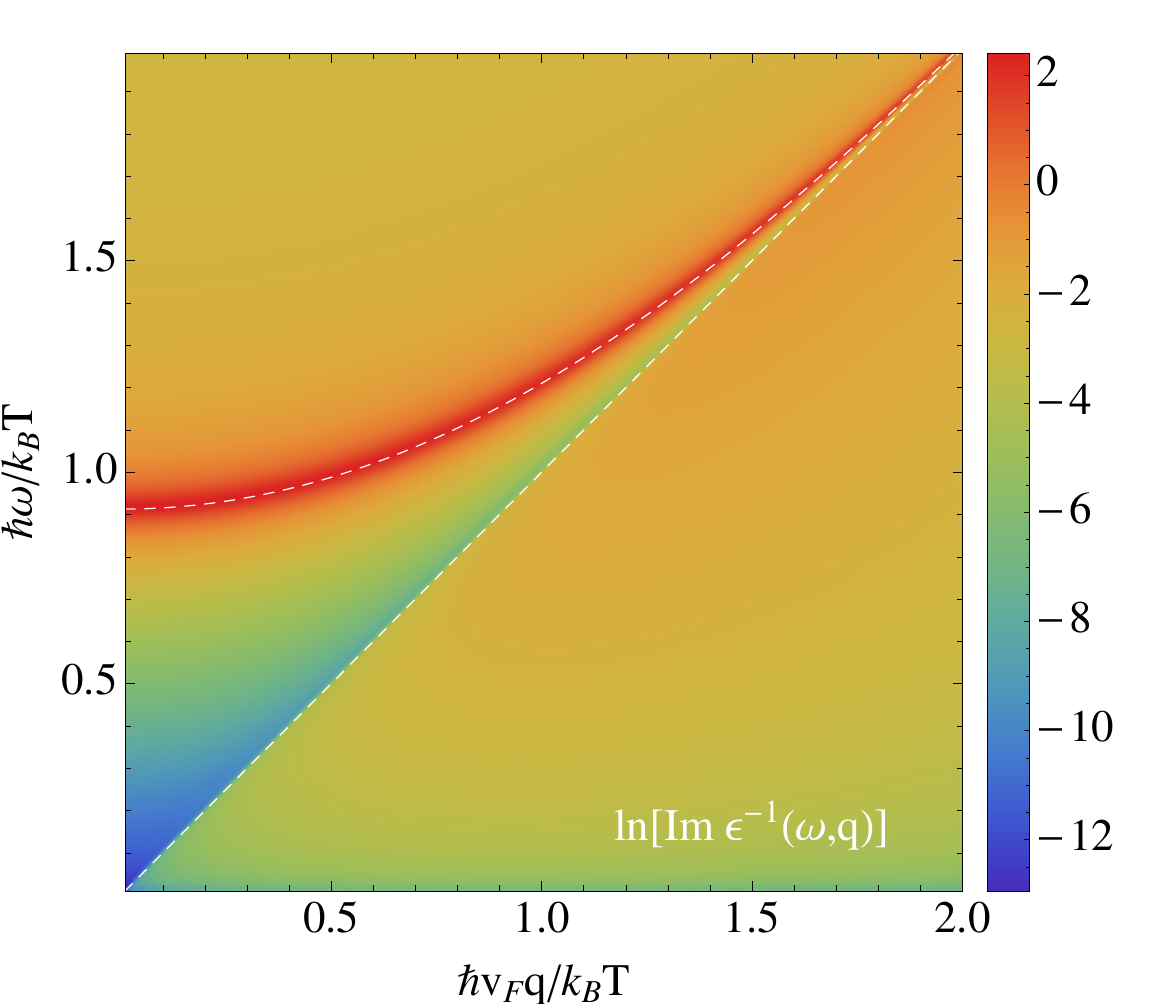}}
\caption{Intrinsic finite temperature dielectric function with $\hbar v_F \Lambda_L/T=5 \cdot 10^3$ ($\Lambda/k_F=100$ and $g \alpha=2.4$). The scale is logarithmic. The white dashed line indicates the plasmon dispersion.}
\label{fig:indiel}
\end{figure}

In the following, we analyze the plasmon dispersion and dielectric function at finite temperature both analytically and numerically. We consider first the intrinsic and then the extrinsic plasmon. For intrinsic Dirac materials, the polarization at small momentum $q$ and high temperature $T$ reads
\begin{align}
\Pi(\omega, q) &= - \frac{g q^2}{12 \pi^2 \hbar v_F} \ln \frac{\hbar v_F \Lambda}{\pi e^{-\gamma_E} T} + \frac{g q^2 T^2}{18 \hbar^3 v_F \omega^2} - \frac{i g q^2 \omega}{96 \pi v_F T} ,
\end{align}
where $\gamma_E$ is the Euler constant, giving rise to the plasmon dispersion
\begin{align}
\hbar \omega_p &= \pi T\sqrt{\frac{2}{3 \ln \hbar v_F \Lambda_L/\pi e^{-\gamma_E} T}} = \sqrt{\dfrac{2 \pi g \alpha}{\kappa(T)}} \, \dfrac{T}{3} . \label{eq:intplasmondispersion}
\end{align}
We have expressed the frequency in terms of the Landau scale $\Lambda_L$ as well as the more conventional form with an effective temperature dependent dielectric constant
\begin{align}
\kappa(T) &= 1 + \frac{g \alpha}{3 \pi} \ln \frac{\hbar v_F \Lambda}{\pi e^{-\gamma_E} T} = \frac{g \alpha}{3 \pi} \ln \frac{\hbar v_F \Lambda_L}{\pi e^{-\gamma_E} T} .
\end{align}
The plasmon temperature dependence is of superlinear form $\omega_p \sim - T/\ln T$, which is one of the main results of this Rapid Communication. For $T=100 \, {\rm K}$ and $\hbar v_F \Lambda_L = 500 \, {\rm K}$ and $5000 \, {\rm K}$, for example, the plasmon energy~\eqref{eq:intplasmondispersion} is $\hbar \omega_p = 20 \, {\rm meV}$ and $10 \, {\rm meV}$, corresponding to a THz frequency range. Note that there is a significant depression of the plasmon frequency compared to a standard Dirac plasma  [i.e., a material without a hole band of negative energy], in which renormalization effects are absent and $\kappa(T)=1$. Both the anomalous temperature dependence and the change in the plasmon frequency we predict here should be readily observable.

Expanding the polarizability to leading order around $\omega_p$, we obtain the damping of the plasmon mode
\begin{align}
\hbar\gamma_p &= \dfrac{\pi^3 T}{24 \ln^2 \hbar v_F \Lambda_L/\pi e^{-\gamma_E}T} = \left(\dfrac{g \alpha}{6 \kappa(T)}\right)^2 \, \dfrac{\pi T}{6} . \label{eq:intplasmondamping}
\end{align}
The plasmon mode is only well defined for $\gamma_p/\omega_p \ll 1$. While both damping and dispersion increase with temperature, it turns out that the ratio
\begin{align}
\frac{\gamma_p}{\omega_p} &= \dfrac{\pi^2}{8 \sqrt{6} \ln^{3/2} \hbar v_F  \Lambda_L/\pi e^{-\gamma_E} T} = \frac{1}{72} \sqrt{\frac{\pi}{2
}} \left(\dfrac{g \alpha}{\kappa(T)}\right)^{3/2} \label{eq:ratio}
\end{align} 
is numerically small. Indeed, the renormalization of the electron charge with $\kappa(T)>1$ decreases this ratio and strongly suppresses the damping relative to the frequency for $T\ll\hbar v_F \Lambda_L$. This is another central result of this work. To further illustrate this point, we show in Fig.~\ref{fig:indiel} the full numerically computed intrinsic dielectric function for a fixed Landau pole scale $\hbar v_F \Lambda_L/T = 5 \cdot 10^3$. It is apparent that density fluctuations are dominated by a sharp plasmon dispersion as discussed above. The plasmon mode at small momentum starts off at constant frequency given by Eq.~\eqref{eq:intplasmondispersion} with a parabolic dispersion, which exists up to large momenta and eventually merges with the particle-hole continuum.

\begin{figure}[t]
\subfigure[]{\scalebox{0.4}{\includegraphics{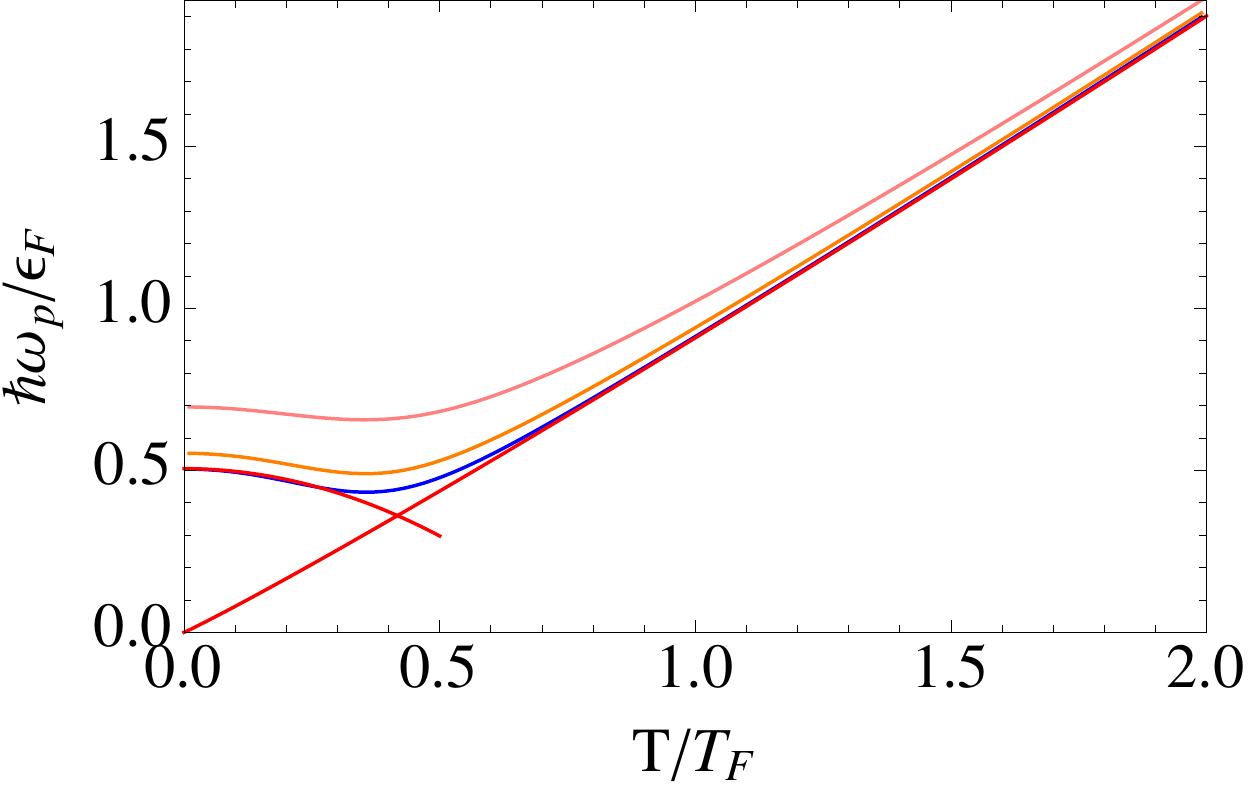}}}
\subfigure[]{\scalebox{0.32}{\includegraphics{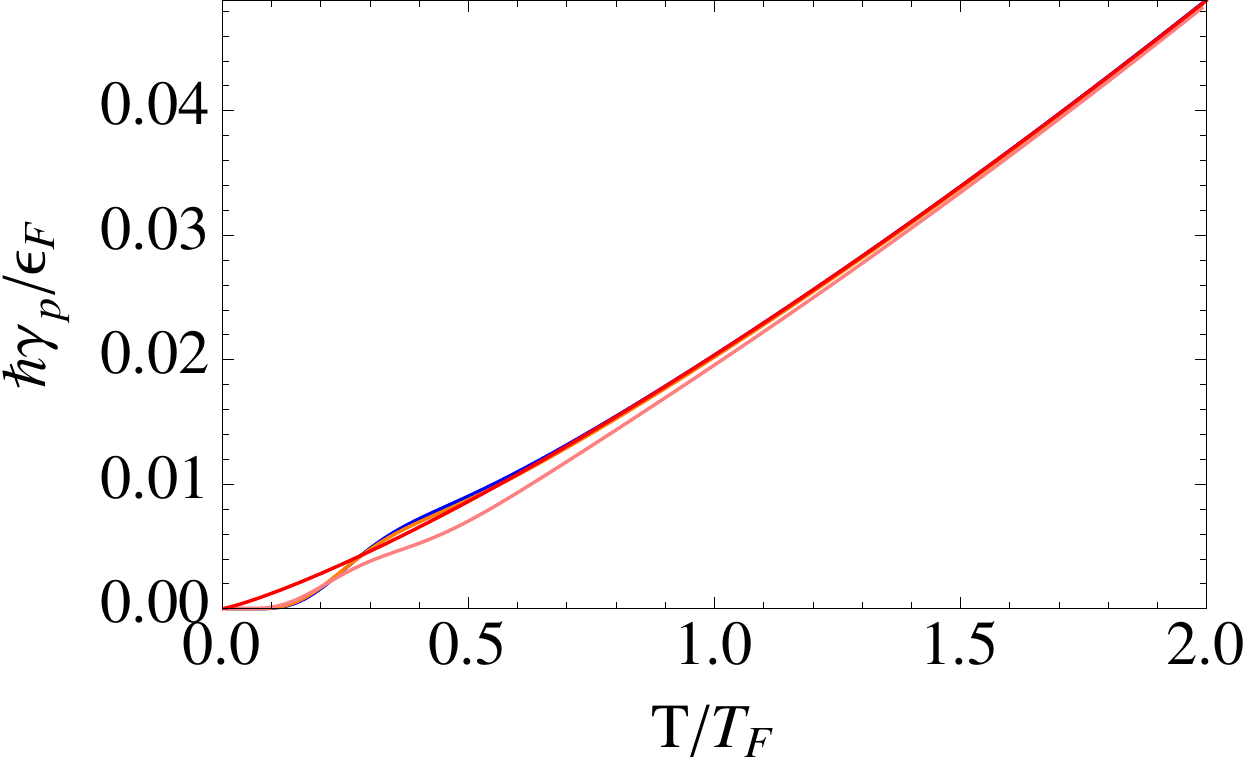}}}
\subfigure[]{\scalebox{0.32}{\includegraphics{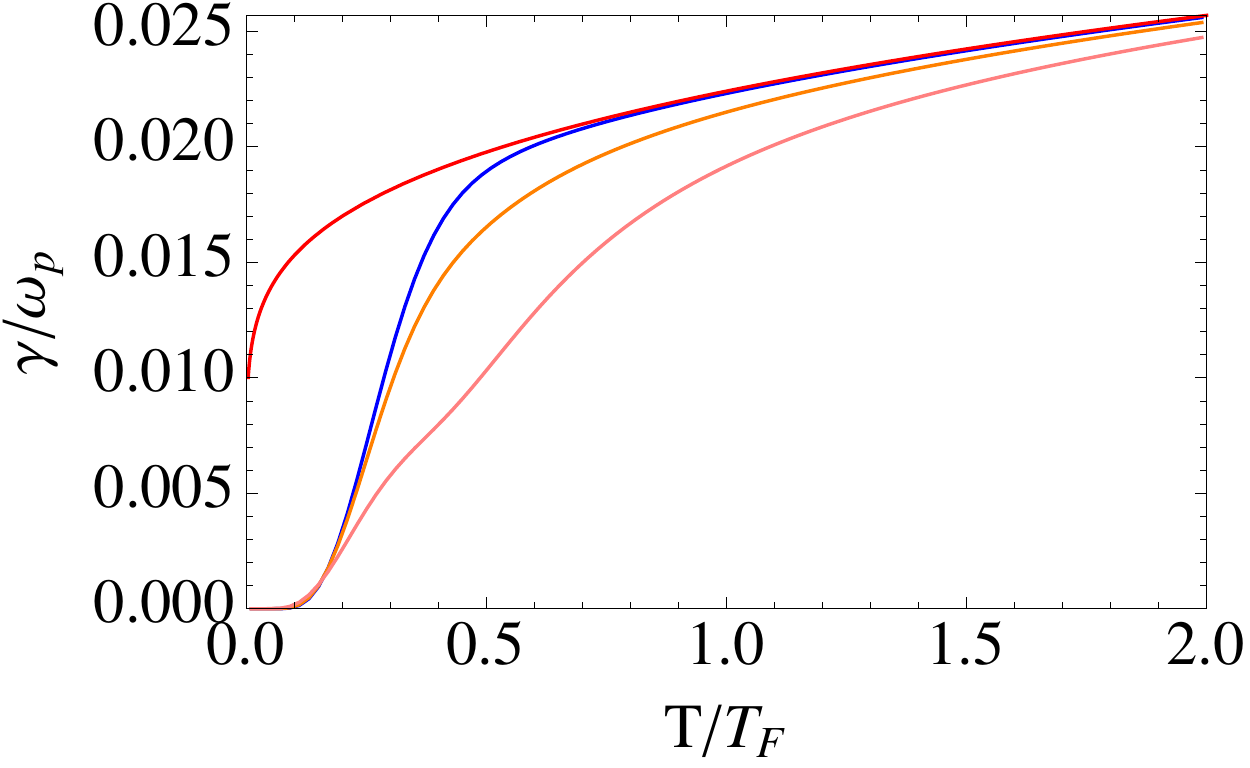}}}
\caption{Extrinsic 3D plasmon dispersion and damping for $g \alpha=2.4$ and $\Lambda/k_F=100$ (corresponding to $\Lambda_L/k_F=5 \cdot 10^3$) for various momenta (bottom to top) $q/k_F=0,0.3,0.6$, and $0.9$. The red continuous lines indicate the exact results derived in Eqs.~\eqref{eq:intplasmondispersion},~\eqref{eq:intplasmondamping},~\eqref{eq:ratio}, and~\eqref{eq:extmulowT}.}
\label{fig:extplasmon}
\end{figure}

The extrinsic plasmon displays a crossover from a genuine extrinsic behavior governed by thermal intraband excitations to the intrinsic behavior at high temperature. In contrast to the intrinsic case, the chemical potential here is nonzero and depends on temperature. It is fixed by the relation
\begin{align}
\frac{g}{V} \sum_{\bf k} n_+({\bf k}) + \frac{g}{V} \sum_{\bf k} [n_-({\bf k}) - 1] &= n ,
\end{align}
where $n_s$ are the Fermi-Dirac distributions with energy $\varepsilon_s(k) = s \hbar v_F k$ and $n=g k_F^3/6 \pi^2$ is the zero temperature carrier density. We find that the chemical potential  has a quadratic deviation from $\varepsilon_F$ at low temperature $\mu/\varepsilon_F = 1 - \pi^2T^2/3T_F^2$, where $T_F$ denotes the Fermi temperature, and goes to zero as a power law at high temperature $\mu/\varepsilon_F = T_F^2/\pi^2T^2$.

\begin{figure*}[t]
\subfigure[$ \ T/T_F=0.01$]{\scalebox{0.27}{\includegraphics{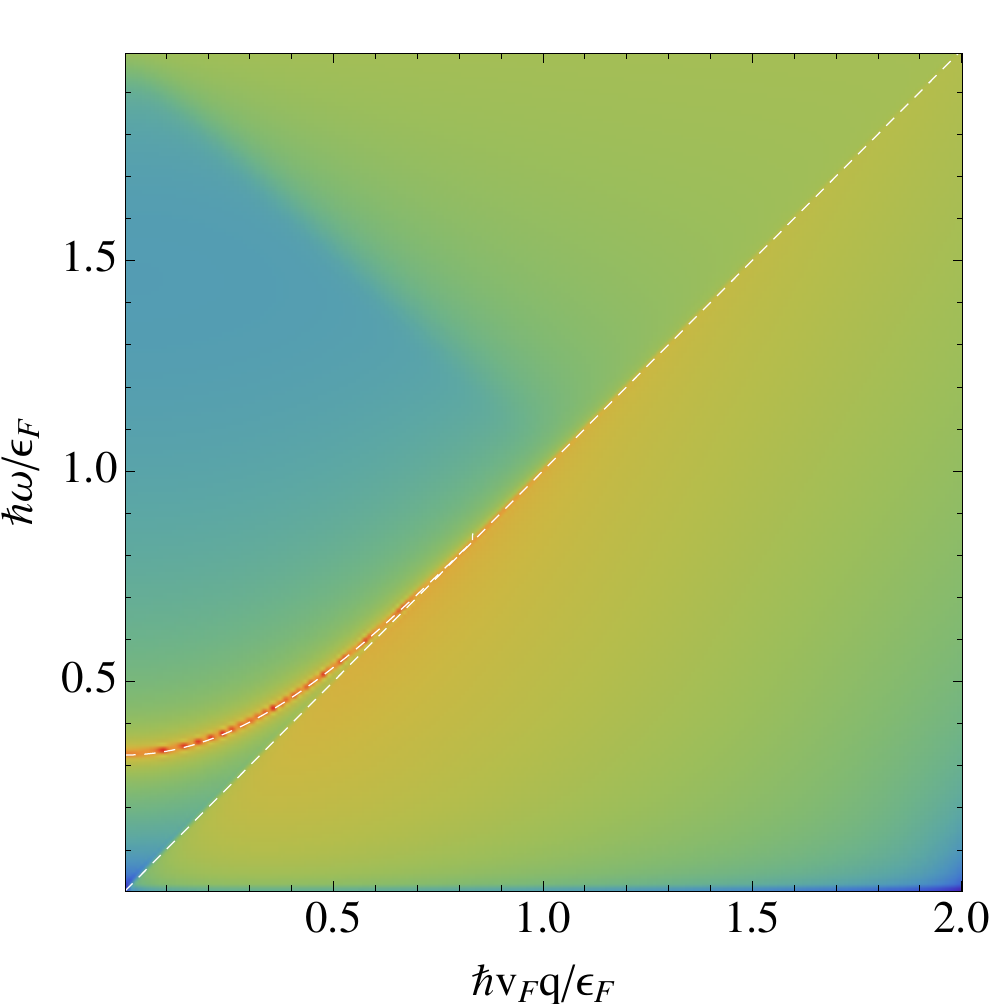}}}
\subfigure[$ \ T/T_F=0.05$]{\scalebox{0.27}{\includegraphics{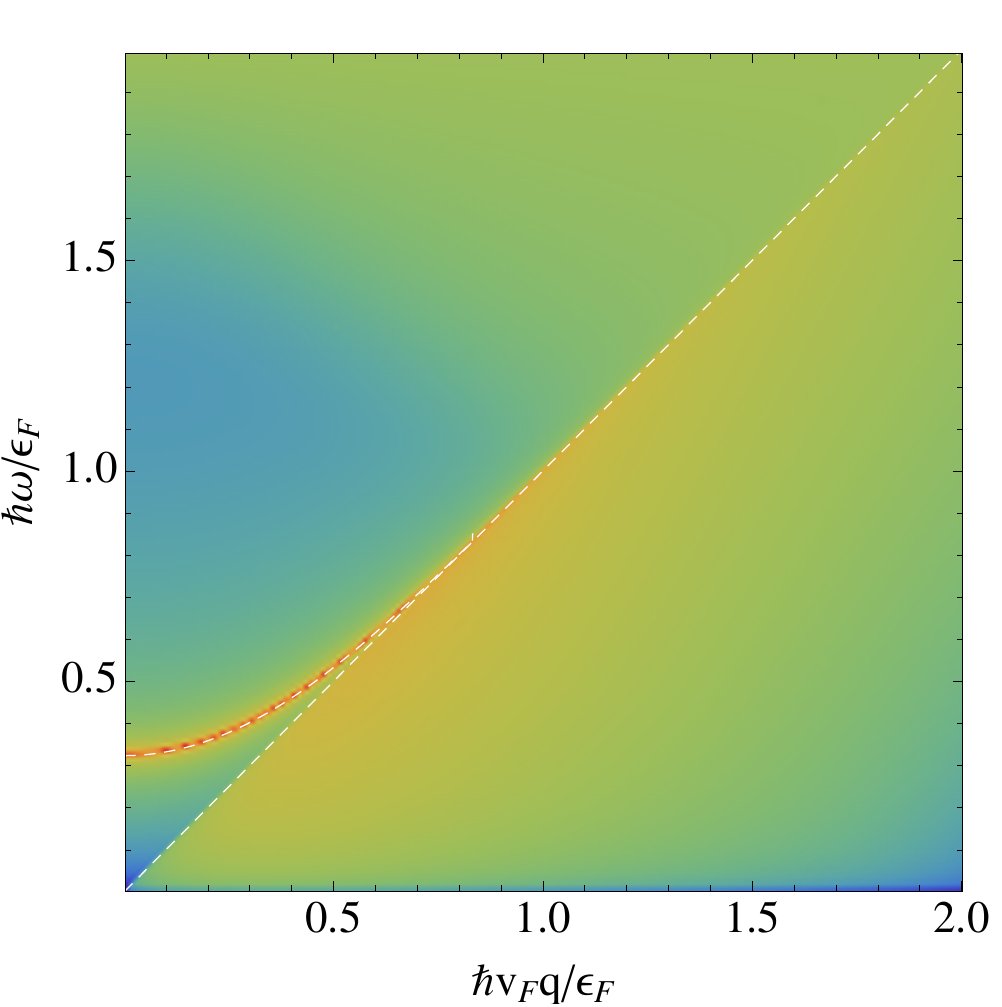}}}
\subfigure[$ \ T/T_F=0.1$]{\scalebox{0.27}{\includegraphics{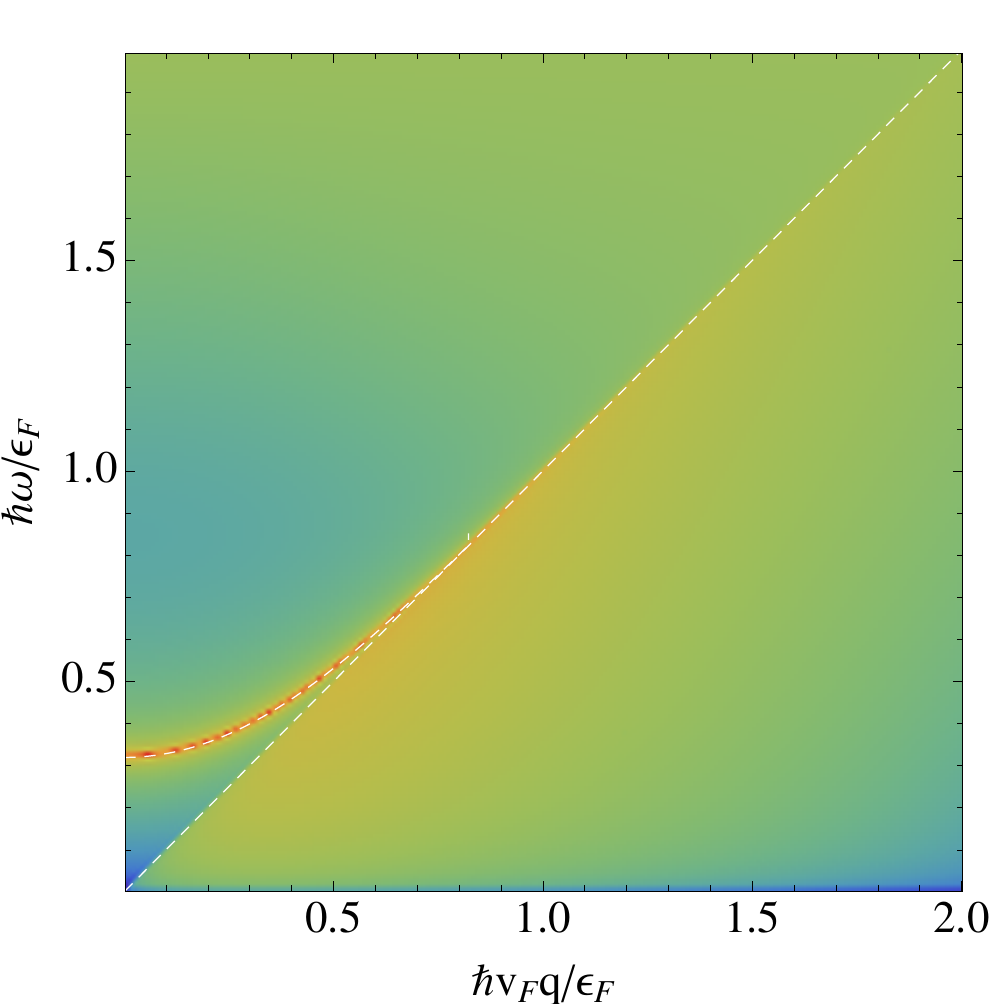}}}
\subfigure[$ \ T/T_F=0.15$]{\scalebox{0.27}{\includegraphics{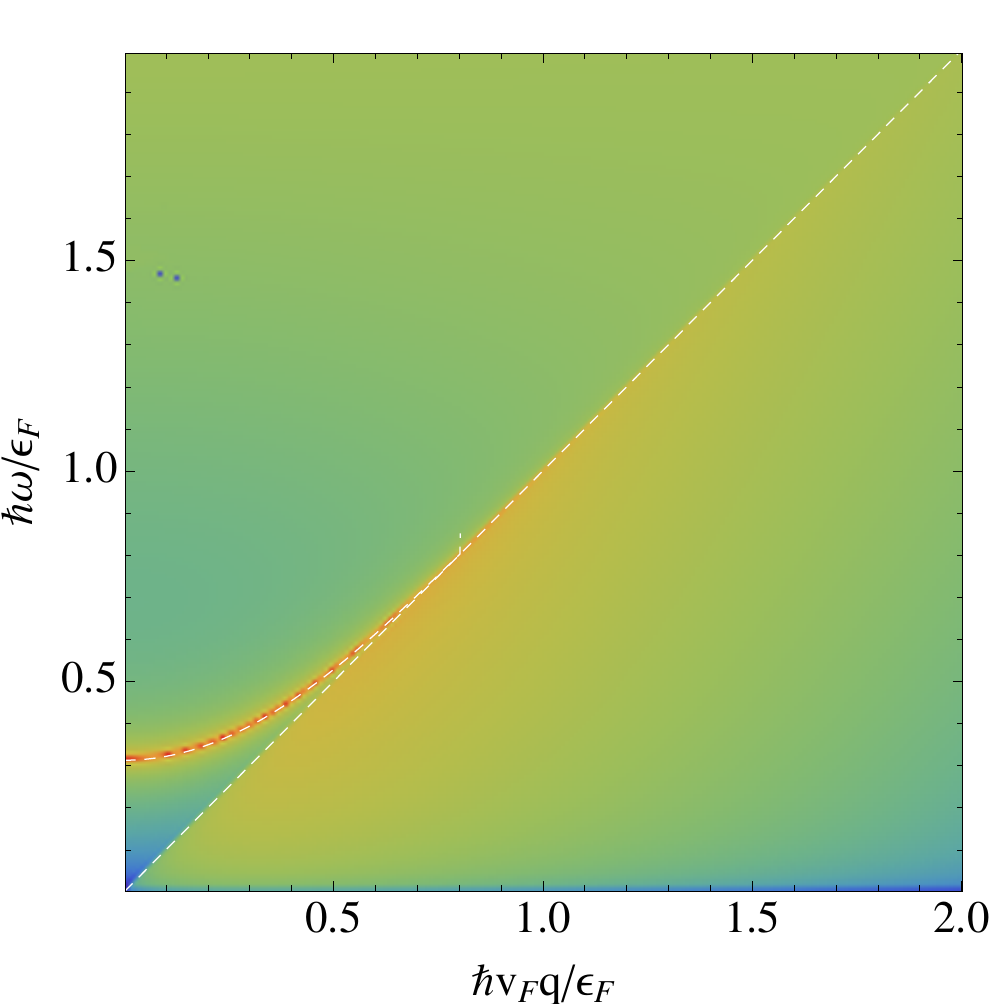}}}
\subfigure[$ \ T/T_F=1.$]{\scalebox{0.27}{\includegraphics{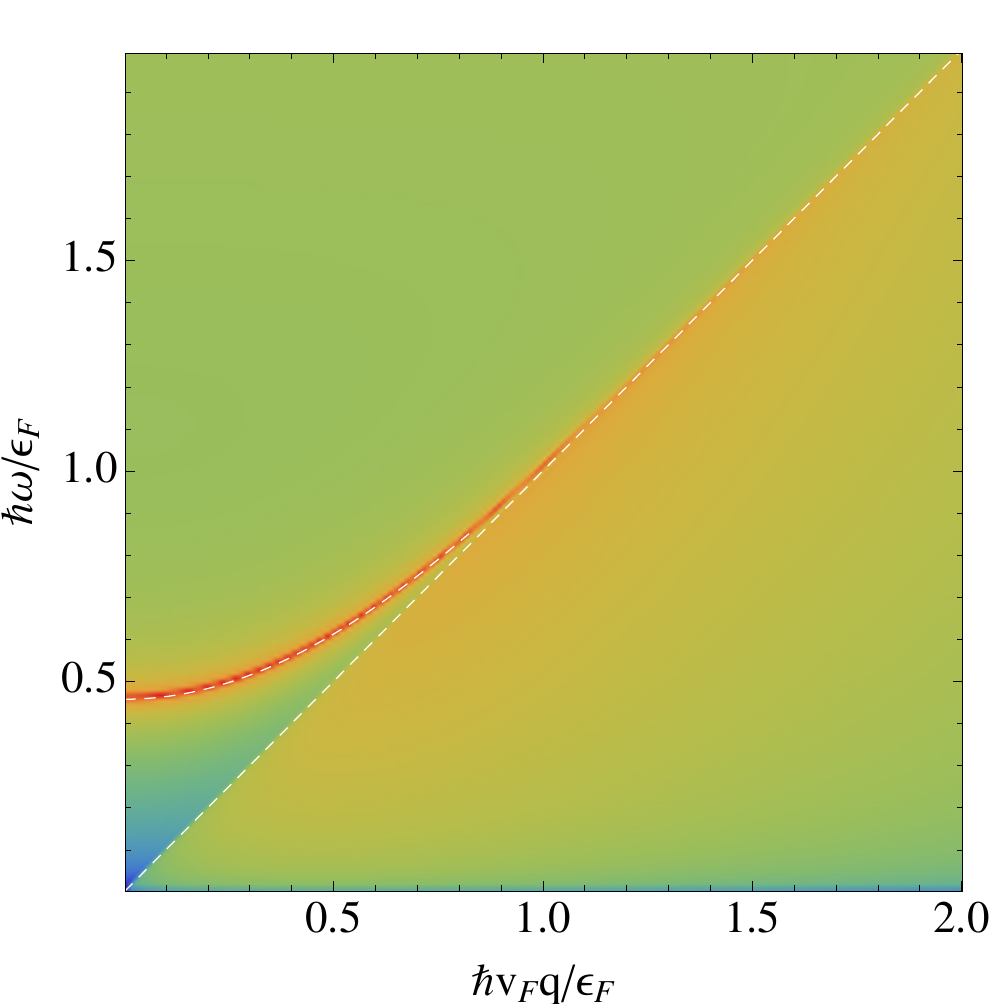}}}
\subfigure[$ \ T/T_F=2.$]{\scalebox{0.27}{\includegraphics{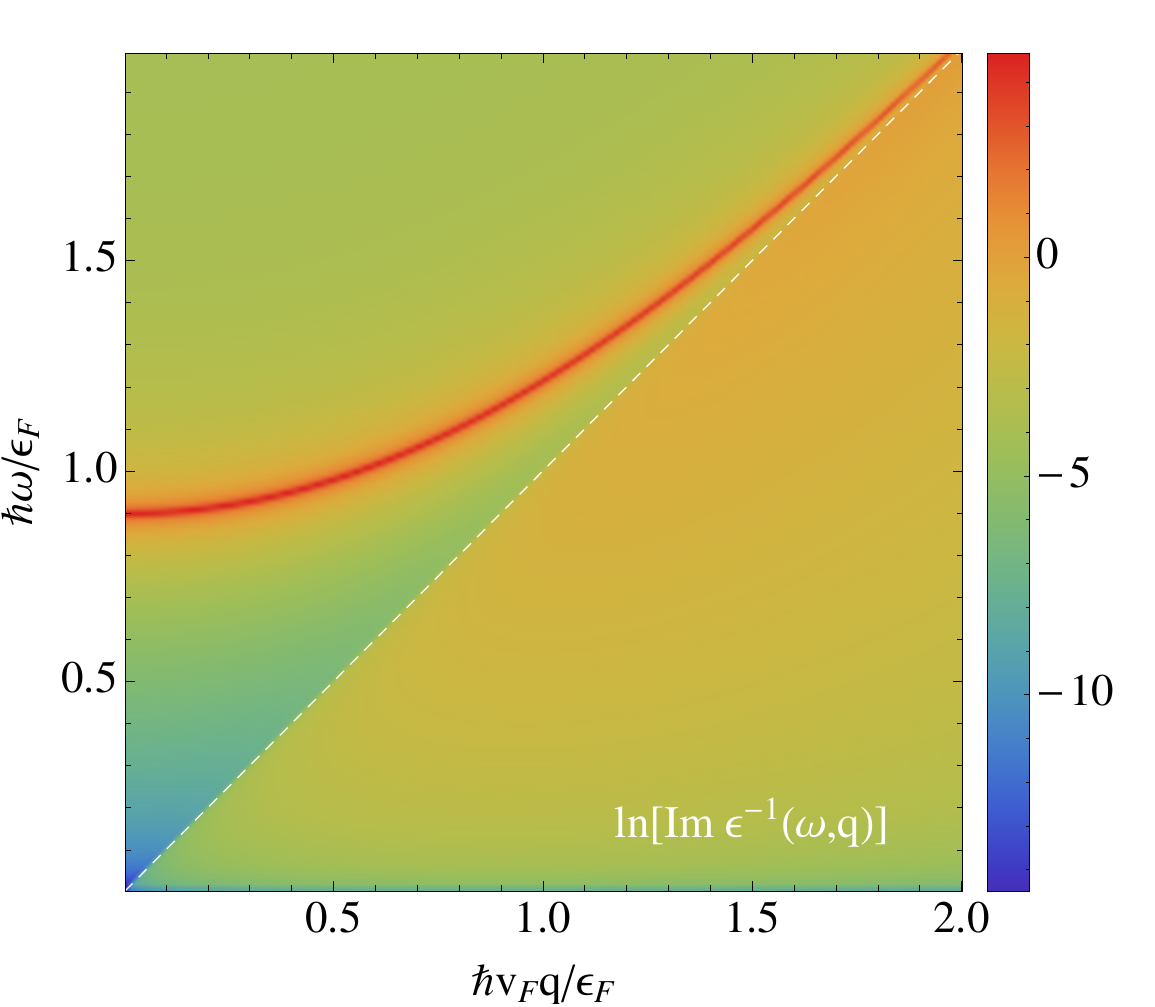}}}
\caption{Extrinsic 3D finite temperature dielectric function for $g \alpha=0.62$ and $\Lambda/k_F=100$ and various temperatures $T/T_F=0.01,0.05,0.1,0.15,1.$, and $2$. The white dashed line indicates the plasmon dispersion. The crossover from the extrinsic zero-temperature dispersion to the intrinsic high-temperature limit is clearly visible.}
\label{fig:extrinsicfinT}
\end{figure*}

At zero temperature, the extrinsic plasmon frequency is given by~\cite{dassarma09,lv13,zhou14}
\begin{align}
\hbar\omega_p^0 &= \sqrt{\frac{2}{\ln \Lambda_L/2k_F}} \varepsilon_F = \sqrt{\dfrac{2 g \alpha}{3 \pi \kappa_0} } \varepsilon_F \sim n^{1/3},
\end{align}
where $\kappa_0$ is a renormalized dielectric constant
\begin{align}
\kappa_0 &= \frac{g\alpha}{3\pi} \ln \frac{\Lambda_L}{2k_F} = 1 + \frac{g \alpha}{3 \pi} \ln \frac{\Lambda}{2 k_F} .
\end{align}
Note that the extrinsic plasmon frequency goes as the cube root of the doping density up to logarithmic corrections. There are two low-temperature corrections to the plasmon dispersion: first, the temperature dependence of the chemical potential discussed in the previous section, and second, a Sommerfeld correction due to the smoothing of the Fermi surface. This gives
\begin{align}
\omega_p(T) &= \omega_p^0
\biggl[1- \dfrac{\pi^2}{6} \biggl(\dfrac{T}{T_F}\biggr)^2\biggr] + {\it O}(T^4/T_F^4) . \label{eq:extmulowT}
\end{align}
The damping of the plasmon mode at low temperatures requires the thermal excitations of particle-hole pairs and is therefore suppressed by powers of $e^{-\mu/T}$.

The intrinsic behavior discussed above also describes the extrinsic material at higher temperatures ${\it O}(T_F)$ as can be seen by computing the high-temperature limit of the extrinsic polarizability,
\begin{align}
\Pi(\omega, q) &= - \frac{g q^2}{12 \pi^2 \hbar v_F} \ln \frac{\hbar v_F \Lambda}{\pi e^{-\gamma_E}  T} + \frac{g q^2 T^2}{6 \pi^2 \hbar^3 v_F \omega^2} \biggl[\frac{\pi^2}{3} + \frac{\mu^2}{T^2}\biggr] \nonumber \\
& - \frac{i g q^2 \omega}{96 \pi v_F T} \biggl[1 - \frac{\hbar^2 \omega^2}{48 T^2} - \dfrac{\mu^2}{4 T^2}\biggr] , \label{eq:polext}
\end{align}
implying that corrections to the intrinsic plasmon dispersion at high temperature are of order ${\it O}(\mu^2/T^2) = {\it O}(1/T^6)$ and thus highly suppressed.

Equations~\eqref{eq:extmulowT} and~\eqref{eq:polext} imply that the plasmon frequency displays a nonmonotonic behavior where it first decreases with increasing temperature manifesting the extrinsic behavior, reaches its minimum, and then increases with the characteristic intrinsic superlinear behavior. The minimum, which is reached at a temperature scale ${\it O}(T_F)$, marks the crossover between genuine extrinsic and intrinsic behavior. The measurement of the plasmon dispersion as a function of temperature thus provides an independent way to determine the extrinsic Fermi energy. The analytic results for the plasmon crossover derived in the previous section are in close agreement with a full numerical calculation of the plasmon dispersion as shown in Fig.~\ref{fig:extplasmon}. The analytical results in Eqs.~\eqref{eq:intplasmondispersion},~\eqref{eq:intplasmondamping},~\eqref{eq:ratio}, and~\eqref{eq:extmulowT} are indicated by red lines. It is apparent that the full result interpolates closely between the low-temperature and the high-temperature exact results. Here, we have the interesting situation of a rapid crossover between the low-temperature regime, where density fluctuations are completely determined by the intraband excitations of the extrinsic system [up to a renormalization of the effective coupling $\kappa(k_F)$], and the high-temperature regime completely characterized by interband excitations. It is important to note that irrespective of the initial detuning, any extrinsic effects will be washed out at sufficiently high temperature. Figure~\ref{fig:extplasmon} also shows the plasmon frequency and damping at nonzero momentum where no analytical results exist. While the damping is almost unchanged in the momentum range considered here, the plasmon dispersion increases with increasing momentum. At high temperature, the intrinsic plasmon dispersion persists (with a higher offset compared to the zero-momentum case), but the minimum is less pronounced and eventually vanishes.

In the following, we discuss the crossover from extrinsic to intrinsic behavior for the full dielectric function. Convenient for our numerical computations is the following integral representation for $\Pi(\omega,{\bf q})$ at finite temperature:
\begin{align}
\Pi(\omega,{\bf q},\mu,T) &= \int_{-\infty}^\infty d\mu' \, \frac{\Pi(\omega,{\bf q},\mu', T=0)}{4 T \cosh^2((\mu'-\mu)/2T)} ,
\end{align}
where the full extrinsic polarizability can be evaluated in closed analytical form at zero temperature~\cite{lv13}. The crossover from extrinsic low-temperature to intrinsic high-temperature behavior is apparent in the full dielectric function as can be seen in Fig.~\ref{fig:extrinsicfinT}. While at zero temperature, interband excitations only exist for frequencies $\omega > 2 k_F - q$, this sharp boundary is quickly washed out even for moderate temperatures. At higher temperatures, the dielectric function is equal to the fully intrinsic result and corresponds to Fig.~\eqref{fig:indiel} up to a rescaling of units.

The discussion so far has focussed on the case of completely symmetric Dirac cones, i.e., with single-particle dispersion $\varepsilon_s(k) = s \hbar v_F k$. By contrast, all experimentally realized Dirac materials (and indeed all theoretically conjectured systems), have a strongly asymmetric structure which is reflected in the particle dispersion $\varepsilon_s(k) = s \hbar \sqrt{v_{Fx}^2 k_x^2 + v_{Fy}^2 k_y^2 + v_{Fz}^2 k_z^2}$ with typically $v_{Fz} \gg v_{Fx}, v_{Fy}$. We note that our discussion fully applies to these systems by a simple replacement of the Fermi velocity $v_F$ by the geometric mean $[v_{Fx} v_{Fy} v_{Fz}]^{1/3}$ and rescaling the momentum ${\bf q}/v_F \to (q_x/v_{Fx}, q_y/v_{Fy}, q_z/v_{Fz})$ in all of our results. We also note that although our results are obtained within RPA, they should be essentially exact for realistic Dirac-Weyl systems where the typical degeneracy factor $g\gg1$ by virtue of the many-valley multiple Dirac cone nature of these systems [$g=24$ in the recently discovered Weyl-semimetal TaAs~\cite{huang15,xu15b}], and all higher-order many-body corrections are suppressed by higher powers of $1/g$.  The logarithmic ultraviolet renormalization is unaffected by higher-order corrections by virtue of the renormalizability of the Coulomb interacting Dirac systems.

We have presented a comprehensive study of collective plasmon modes in 3D Dirac materials at finite density and temperature. We find that the plasmon mode is extremely well defined with damping rates more than an order of magnitude smaller compared to the energy. Moreover, the plasmon dispersion is superlinear in temperature with logarithmic Landau pole corrections due to the renormalization of the electron charge. Our results are of fundamental interest as a manifestation of charge renormalization similar to the one found in quantum electrodynamics, and of significant practical relevance as temperature offers a way to tune the plasmon dispersion at will. Our predictions should be directly observable in present-day experiments and should provide definitive signatures for the Dirac nature of candidate electron systems.

\emph{Note added.} Recently, Ref.~\cite{kharzeev14} appeared, which studies intrinsic finite-temperature plasmons in Dirac semimetals and reports the intrinsic plasmon dispersion. Our result reported in Eq.~\eqref{eq:intplasmondispersion} differs from Ref.~\cite{kharzeev14} by a factor of $\pi e^{-\gamma_E}$ and a different definition of the Coulomb interaction, which includes a dielectric constant $\kappa$.

This work is supported by LPS-MPO-CMTC.

\bibliography{bib}

\end{document}